\documentclass[10pt]{amsart}

\newcommand{\limfunc}[1]{\operatornamewithlimits{#1}}

\begin{document}

\title{Security of the Cao-Li Public Key Cryptosystem}
\author[L.-H.~Lim]{Lek-Heng Lim}
\address{DSO National Laboratories, 25th Storey, Tower A, Defence
Technology Towers, Depot Road, Singapore 109679}
\curraddr{Department of Mathematics, Malott Hall, Cornell University,
Ithaca, NY, 14853-4201}
\email{lekheng@math.cornell.edu}

\begin{abstract}
We show that the Cao-Li cryptosystem proposed in \cite{CL1} is not secure.
Its private key can be reconstructed from its public key using elementary
means such as \emph{LU}-decomposition and Euclidean algorithm.
\end{abstract}
\maketitle

\section{Description of the Cryptosystem}

The Cao-Li public key cryptosystem was first proposed in \cite{CL1}. It
encrypts messages using a bilinear form that is chosen to permit easy
decryption by the Chinese remainder theorem. Public key cryptosystems that
are designed along this line are not uncommon in the Chinese cryptographic
literature. However, as most of the original papers were published in
Chinese, they remained relatively obscure until a few of them were described
in \cite{DPS1} (in English) recently. Our description below is based on the
latter reference.

Let $p_1, \dots, p_n$ be $n$ distinct primes where $p_i \equiv 3 
\pmod{4}$. For $i = 1, \dots, n$, define 
\[
m_i := \frac{1}{p_i} \Biggl(\prod_{j=1}^n p_j \Biggr)
\text{.} 
\]
Compute for each $m_i$, an integer $m_i'$ that satisfies $m_i' m_i \equiv
1 \pmod{p_i}$ and $0 < m_i' < p_i$. We define positive integers 
\[
\lambda_i := m_i' m_i 
\]
for $i = 1, \dots, n$ and the diagonal matrix 
\[
\Lambda := \limfunc{diag}\left[ \lambda_1, \dots, \lambda_n \right] . 
\]
Note that 
\begin{equation}
\lambda_i \equiv \delta_{ij} \pmod{p_j} \label{a}
\end{equation}
where $\delta_{ij}$ is $1$ if $i = j$ and $0$ otherwise.

We choose another two invertible $n \times n$ lower-triangular matrices
$P_1$ and $P_2$ with non-negative integer entries that are bounded by 
\begin{equation}
\beta := \min_{1 \le i \le n} \sqrt{\frac{p_i}{i(i + 1)d}} \label{b}
\end{equation}
where $d \ge 1$ is a chosen positive integer.

The secret key comprises the two matrices $P_1, P_2$ and the primes $p_i$, 
$i = 1, \dots, n$. The public key is the $n \times n$ symmetric matrix $B$
given by 
\[
B := P_2^T P_1^T \Lambda P_1 P_2 \text{.} 
\]
Let the message block be $\mathbf{x} = (x_1, \dots, x_n)$ where $0 \le
x_i \le d$. The ciphertext $y$ is computed as 
\[
y = \mathbf{x} B \mathbf{x}^T \text{.} 
\]
If we let $\mathbf{z} := \mathbf{x} P_2^T P_1^T$, then 
\[
y = \mathbf{z} \Lambda \mathbf{z}^T = \lambda_1 z_1^2 + \dots + \lambda_n 
z_n^2 \text{.} 
\]
From \eqref{a}, we have 
\begin{equation}
z_k^2 \equiv y \pmod{p_k} \text{.}  \label{c}
\end{equation}
Keeping in mind that $P_1^T$ and $P_2^T$ are upper-triangular and their
entries are non-negative and bounded by $\beta$, we have, from \eqref{b}
and $0 \le x_i \le d$, that 
\begin{equation}
0 \le z_k \le \sum_{i=1}^k \sum_{j=i}^k d\beta^2 = d\beta^2 
\frac{k(k+1)}{2} < \frac{p_k}{2} \text{.}  \label{d}
\end{equation}
We can carry out decryption as follows. For each $k = 1, \dots, n$,
compute the unique $z_k$ satisfying \eqref{c} and \eqref{d}. The message
can then be recovered by 
\begin{equation}
\mathbf{x} = \mathbf{z} \left( P_2^T P_1^T \right)^{-1} \text{.}
\label{e}
\end{equation}
Note that since $p_k \equiv 3 \pmod{4}$, effective algorithms for
computing square roots $\pmod{p_k}$ exist (see \cite{C1}).

\section{Key Recovery}

We will first recover $\Lambda$ from $B$. Let $P_1P_2 =: P = 
(p_{ij})_{1 \le i, j \le n}$. Then $P$ is an invertible lower-triangular
matrix with non-negative integral entries by the same properties of $P_1$
and $P_2$. Since $P$ is invertible and has non-negative integral entries,
we have $\det P = 1$. Moreover, we also have $\det P = p_{11} \times \dots
\times p_{nn}$ since $P$ is triangular. As all the $p_{ii}$'s are
non-negative, it then follows that $p_{ii} = 1$ for $i = 1, \dots, n$.

$\Lambda$ and $P$ can be recovered from $B$ using an algorithm that is very
similar to the algorithm for \textit{LU}-decomposition of a matrix (the
difference being that row reduction is done starting from the bottom rows).
Denote the $i$th row of $B$ by $\mathbf{b}_i = (b_{i1}, \dots, b_{in})$,
$i = 1, \dots ,n$. We know immediately that $b_{nn} = \lambda_n$.
\begin{table}[htbp]
\centering
\leavevmode
\begin{tabular}{|lll|}
\hline
\multicolumn{3}{|l|}{\textbf{Algorithm A}}
\\ \hline\hline
\textsc{Input.} &  & $ B = (\mathbf{b}_1, \dots, \mathbf{b}_n)^T =
(b_{ij})_{1 \le i,j\le n}$
\\ \hline
\textsc{Output.} &  & $\lambda _1,\dots ,\lambda _n,P$
\\ \hline
\textsc{Step 1.} &  & \textsf{for} $i=n-1,n-2,\dots ,1$ \textsf{do} \\ 
&  & \qquad \textsf{for} $j=n,n-1,\dots ,i+1$ \textsf{do} \\ 
&  & \qquad \qquad $\mathbf{b}_i\leftarrow \mathbf{b}_i-\dfrac{b_{ji}}{b_{jj}%
}\mathbf{b}_j$\textsf{;} \\ 
&  & \qquad \textsf{end;} \\ 
&  & \textsf{end;} \\ \hline
\textsc{Step 2.} &  & \textsf{for} $i=1,\dots ,n$ \textsf{do} \\ 
&  & \qquad $\lambda _i\leftarrow b_{ii}$\textsf{;} \\ 
&  & \qquad $\mathbf{b}_i\leftarrow \mathbf{b}_i/\lambda _i$\textsf{;} \\ 
&  & \textsf{end;} \\ 
&  & $P\leftarrow B$\textsf{;} \\ \hline
\end{tabular}
\end{table}

The following shows that Algorithm A indeed yields the required output. Let
the $i$th row of $P$ be $\mathbf{p}_i$, $i = 1, \dots, n$. Since 
$p_{ji} = 0$ if $j < i$ and $p_{ii} = 1$, we may write $\mathbf{b}_i =
\lambda_i \mathbf{p}_i + \sum_{j = i + 1}^n \lambda_j p_{ji} 
\mathbf{p}_j$. For each $i = n - 1, n - 2, \dots, 1$, the inner loop of
\textsc{Step 1} effectively does
\[
\mathbf{b}_i \leftarrow \mathbf{b}_i - \sum_{j = i + 1}^n 
\dfrac{b_{ji}}{b_{jj}} \mathbf{b}_j \text{.}
\]
We shall show inductively that $\mathbf{b}_i$ is reduced to $\lambda_i 
\mathbf{p}_i$ at stage $i$: clearly $\mathbf{b}_n = \lambda_n 
\mathbf{p}_n$; suppose $\mathbf{b}_i$ is reduced to $\lambda_i 
\mathbf{p}_i$ at stage $i = n - 1, \dots, n - k$, then at stage $n - k 
- 1$, 
\begin{align*}
\mathbf{b}_{n - k - 1}
&\leftarrow \mathbf{b}_{n - k - 1} - \sum_{j = n - k}^n 
\dfrac{b_{ji}}{b_{jj}} \mathbf{b}_j \\
&= \mathbf{b}_{n - k - 1} - \sum_{j = n - k}^n \lambda_j p_{ji}
\mathbf{p}_j \\
&= \lambda_{n - k - 1} \mathbf{p}_{n - k - 1}\text{.}
\end{align*}
Hence \textsc{Step 1} reduces $B = (\mathbf{b}_1, \dots, \mathbf{b}_n)^T$
to $(\lambda_n \mathbf{p}_n, \dots, \lambda_n \mathbf{p}_n)^T = \Lambda
P$. Since the diagonal entries of $P$ are all $1$'s, the diagonal entries
of $\Lambda P$ are the required $\lambda_i$'s. Consequently, $P$ can be
recovered by dividing each row by its corresponding diagonal entry.

We can now recover the moduli $p_1, \dots ,p_n$ from $\lambda_1, \dots, 
\lambda_n$. From \eqref{a}, we see that for a fixed $i$, $p_i \mid
\lambda_j$ for all $j \neq i$ and $p_i \mid \lambda_i - 1$. So 
\[
p_i \mid d_i := \gcd (\lambda_1, \dots, \lambda_{i - 1}, \lambda_i - 1,
\lambda_{i + 1}, \dots, \lambda_n)\text{.} 
\]
It could of course happen that $d_i \neq p_i$ for some $i$. So this
process only partially recovers the $p_i$'s. However our computer
simulations (using C++ with \textsf{LiDIA}) show that instances where
$d_i \neq p_i$ are rare. We shall give some heuristics to substantiate
this claim. For $d_i = p_i$, it is sufficient that $\gcd (m_1', \dots,
m_{i - 1}', m_{i + 1}', \dots, m_n') = 1$. From \cite{HST1}, we have 
\begin{gather*}
\#\bigl\{ (a_1, \dots, a_k) \in \mathbb{N}^k \bigm| \gcd (a_1, \dots,
a_k) = 1, \text{ all } a_i \le N \bigr\} \\
= \begin{cases}
N^k/\zeta(k) + \limfunc{O}(N^{k - 1}) &\text{if }k > 2, \\ 
6N^2/\pi^2 + \limfunc{O}(N\log N)     &\text{if }k = 2.
\end{cases}
\end{gather*}
where $\zeta (s) = \sum_{i = 1}^\infty i^{-s}$ is the Riemann zeta
function. Assuming that each $m_i'$ is randomly distributed in $\{1, 
\dots, N \}$ where $N := \max \{p_1, \dots,p_n\}$, the probability that
$\gcd (m_1', \dots, m_{i - 1}', m_{i + 1}', \dots, m_n') = 1$ is then at
least $\zeta(n-1) \ge 6/\pi^2 \approx 0.60$ when $N$ is large enough. So
we can expect to recover more than half of the $p_i$'s. In fact our
simulations show that we almost always have $d_i = p_i$ and many of the
rare exceptions are of the form $d_i = 2p_i$ where $p_i$ can also be
recovered easily.

\section{Conclusion}

Note that Algorithm A is essentially \textit{LU}-decomposition and the
$d_i$'s can be computed using the Euclidean algorithm. Since these two
methods can be carried out efficiently, we can easily recover $P$ and most
of the $p_i$'s. It then follows that the Cao-Li cryptosystem is insecure
and thus should not be used.

\end{document}